\documentclass[superscriptaddress,showpacs,showkeys,floatfix,number,review]{elsarticle}
\usepackage{graphicx}
\usepackage{amssymb}
\journal{Physics Letters A}
\begin{document}
\begin{frontmatter}

\title{The Planck era with a negative cosmological constant and cosmic strings}

\author[rvt]{G. A. Monerat}

\author[rvt]{L. G. Ferreira Filho}

\author[els]{G. Oliveira-Neto\corref{cor1}\fnref{fn1}}
\ead{gilneto@fisica.ufjf.br}

\author[rvt]{E. V. Corr\^{e}a Silva}

\author[rvt]{C. Neves}

\address[rvt]{Departamento de Matem\'{a}tica, F\'{\i}sica e Computa\c{c}\~{a}o,
           Faculdade de Tecnologia,
           Universidade do Estado do Rio de Janeiro,
           Rodovia Presidente Dutra Km 298, P\'{o}lo Industrial,
           CEP 27537-000, Resende, R. J., Brazil.}

\address[els]{Departamento de F\'{\i}sica, 
              Instituto de Ci\^{e}ncias Exatas,
              Universidade Federal de Juiz de Fora,
              CEP 36036-330, Juiz de Fora, M. G., Brazil.}          

\cortext[cor1]{Corresponding author}  

\fntext[fn1]{phone/fax 55-32-2102-3307(ext. 223)/55-32-2102-3312}      

\begin{abstract}
In the present letter, we consider the {\it DeBroglie-Bohm} interpretation 
of quantum Friedmann-Robertson-Walker (FRW) models in the presence of a 
negative cosmological constant and cosmic strings. We compute the Bohm's
trajectories and quantum potentials for a quantity related to the scale 
factor. Then, we compare our results with the ones already in the literature,
where the many worlds interpretation of the same models was used.
\end{abstract}


\end{frontmatter}

Quantum cosmology was the first attempt in order to remove the Big
Bang singularity by quantizing the gravitational theory \cite{dewitt}.
The {\it DeBroglie-Bohm} interpretation of quantum mechanics \cite{bohm}, 
\cite{holland}, is frequently used in quantum cosmology. In the 
minisuperspace models treated using the {\it DeBroglie-Bohm} 
interpretation the common argument used to justify the absence of a Big Bang 
singularity is the fact that the scale factor Bohmian trajectories $a(t)$, as 
a function of a chosen time, never go through $a=0$ 
\cite{germano, germano1, acacio, acacio0, gil}.

Many authors have quantized FRW models with different kinds of perfect fluids 
\cite{germano, pedram, nivaldo, nivaldo2, gotay, nelson, rubakov, barboza} 
using the variational formalism developed by Schutz \cite{schutz}. An
important example of a perfect fluid is the one representing cosmic strings.
Cosmic strings were originally thought as linear defects formed at symmetry 
breaking phase transitions in the early universe. They would play a fundamental
role in the structure formation of our universe. Unfortunately, due to some of
their properties and observational bounds it became clear that cosmic strings 
could not produce any relevant effect on the issue of structure formation 
\cite{polchinski}. More recently, it was demonstrated that in models where the
universe has extra dimensions the fundamental superstrings may have cosmic length 
\cite{polchinski}. Since they have different properties from the original cosmic
strings, described above, they may still play an important hole in structure
formation. Apart from that, any kind of cosmic string may produce some observational
effects. The most important of them are gravitational waves and cosmic lensing 
\cite{vilenkin}.

In this work we consider the {\it DeBroglie-Bohm} interpretation of quantum FRW 
models in the presence of a negative cosmological constant and cosmic strings.
Each model differs from the others due to the curvature of the spatial sections
which may be positive, negative or null. Cosmic strings may be described by a 
perfect fluid with equation of state $p=-(1/3)\rho$, where $p$ is the fluid 
pressure and $\rho$ is its density. The presence of a negative cosmological
constant in the present models implies that the universes described by them
have maximum sizes, in other words they are bounded. Although recent observations 
point toward a positive cosmological constant, it is still possible that at the 
very early Universe the cosmological constant was negative.
Those models have already been quantized using
the variational formalism developed by Schutz \cite{pedram}. There, the authors
used the spectral method to compute the energy eigenvalues and eigenfunctions 
of the Wheeler-DeWitt equations corresponding to each model. Then, they combined
the energy eigenfunctions and the time dependent sector in order to obtain
wave packets. From these wave packets they derived the time-dependent scale factor
expectation values for each model. For each model, it is clear that the scale
factor expected values never go to zero. These results give a strong indication 
that those models are free from singularities, at the quantum level. 

In order to apply the {\it DeBroglie-Bohm} interpretation for those models, it 
would be easier if we had algebraic expressions for each wavefunction. We obtain
those expressions by first applying the canonical transformation introduced in 
Refs. \cite{barboza} and \cite{gil1}, to the superhamiltonian eq. (8) of 
Ref. \cite{pedram}. In the present situation this equation reduces to,
\begin{equation}
\label{1}
\mathcal{H}=-\frac{p_{a}^2}{12a^2}-3k+\Lambda a^{2}+p_{T}=0,
\end{equation}
where $a$ and $T$ are the canonical variables and $p_a$ and $p_T$ are their
canonically conjugated momenta, respectively. Using the following canonical
transformation \cite{barboza} and \cite{gil1},
\begin{equation}
\label{2}
a= \sqrt{2x},\,\,\, p_a = p_x a,
\end{equation}
the superhamiltonian function eq. (\ref{1}) transforms to,
\begin{equation}
\label{3}
H=-\frac{p_{x}^2}{12}-(3k-2\Lambda x)+p_{T}=0.
\end{equation}
It is important to notice that the new set of canonical variables and 
corresponding canonically conjugated momenta are $(x, p_x)$ and $(T, p_T)$.
Also, the physical properties of the models are not modified by the
transformation eq. (\ref{2}) because the superhamiltonians $\mathcal{H}$ 
eq. (\ref{1}) and $H$ eq. (\ref{3}) are identical up to a constant \cite{jose}.
It means that one may describe equally well the models by using the variables 
$(a, T)$ or $(x, T)$.

Now, the quantum dynamics is ruled by the Wheeler-DeWitt 
equation, 
\begin{equation}
\label{4}
\left(-\frac{1}{12}\frac{\partial^2}{\partial x^2}+
(3k - 2\Lambda x)\right)\Psi(x,t)=
i\frac{\partial \Psi(x,t)}{\partial t}
\end{equation}
In order to solve it we assume that,
\begin{equation}
\label{5}
\Psi(x,t) =\eta(x)\exp{\left(-iE_{n}t\right)}.
\end{equation}
It gives rise to the following equation to $\eta (x)$,
\begin{equation}
\label{6}
-\frac{1}{12}\frac{d^2}{dx^2}\eta(x) + 
\left(3k-2\Lambda x-E_{n}\right)\eta(x)=0.
\end{equation}
The solutions to this equation are the Airy functions, algebraic 
expressions as we wished,
\begin{displaymath}
\label{7}
\eta(x) = c_{1} Ai\left(
\frac{\sqrt[3]{3}\left(3k-2\Lambda x-E_{n}\right)}{\Lambda^{2/3}}\right)+
c_{2} Bi\left(
\frac{\sqrt[3]{3}\left(3k-2\Lambda x-E_{n}\right)}{\Lambda^{2/3}}\right).
\end{displaymath}
The Airy functions $Bi$ grow up exponentially when $x\rightarrow \infty$. 
In order to eliminate this undesirable behavior, we put $c_{2}=0$. 
Then, the energy eigenfunctions for our models are,
\begin{equation}
\label{8}
\eta(x)= c_{1} Ai\left(
\frac{\sqrt[3]{3}\left(3k-2\Lambda x-E_{n}\right)}{\Lambda^{2/3}}\right).
\end{equation}
The wave packets can be constructed via the superposition of the energy 
eigenfunctions and the time dependent sector for a given $\Lambda$ value. 
Here, we consider the eigenfunctions and the time dependent sector 
corresponding to the lowest 21 energy levels, in analogy with Ref. \cite{pedram}:
\begin{equation}
\Psi(x,t)= \sum_{n=0}^{20} C_n Ai\left(
\frac{\sqrt[3]{3}\left(3k-2\Lambda x-
E_{n}\right)}{\Lambda^{2/3}}\right)\exp{(-iE_{n}t)}.
\label{9}
\end{equation}

These packets must be identically null at $x=0$. Thus
\begin{equation}
Ai\left(
\frac{\sqrt[3]{3}\left(3k-E_{n}\right)}{\Lambda^{2/3}}\right) = 0.
\label{10}
\end{equation}
The Airy functions $Ai$ have many nodes. This implies that there 
will be a certain discrete set of values of $E_{n}$, obtained as 
solutions to eq. (\ref{10}). As an example, the values of $E_{n}$ for
the cases $k=-1,0,1$ and $\Lambda=-15$ are very similar to the ones
found in Ref. \cite{pedram}, specially for large values of $n$.
The case $\Lambda=0$ recovers the results obtained in Ref. \cite{germano}. 
We restrict ourselves to the cases where $\Lambda<0$. 


The time evolution of the wave packets built from eq. (\ref{9}), for all values 
of $k$, shows that they are null not only at the origin but they are asymptotically 
null at infinity as well. In the region near $x=0$ these packets present strong 
oscillations, which decrease as $x$ increases. 


In order to use the {\it DeBroglie-Bohm} interpretation we must
re-write $\Psi (x,t)$ eq. (\ref{9}), in the polar form,
\begin{equation}
\Psi(x,t)=R(x,t) e^{iS(x,t)}
\label{12}
\end{equation}
where,
\begin{equation}
R(x,t)=
\sqrt{\sum_{n,m=0}^{N} C_n C_m Ai\left( G \left( k,\Lambda,E_n,x \right)\right)
Ai\left( G \left( k,\Lambda,E_m,x \right)\right)\cos\left((E_{n}-E_{m})t\right)}
\label{13}
\end{equation}

\begin{equation}
S(x,t)=\arctan\left[
\frac{-\sum_{n=0}^{N} C_n Ai\left( G \left( k,\Lambda,E_n,x \right)\right)\sin(E_{n}t)}
{\sum_{m=0}^{N} C_m Ai\left( G \left( k,\Lambda,E_m,x \right)\right)\cos(E_{m}t)}
\right].
\label{14}
\end{equation}

\noindent
where $G \left( k,\Lambda,E_n,x \right)=
\frac{\sqrt[3]{3}\left(3k-2\Lambda x-E_{n}\right)}{\Lambda^{2/3}}$ and $N=20$.

Following the {\it Bohm-deBroglie} interpretation we introduce 
$\Psi (x,t)$ in eq. (\ref{4}), this leads to the next two
equations for $R (x,t)$ and $S(x,t)$ \cite{holland},

\begin{eqnarray}
\label{15}
\frac{\partial S(x,t)}{\partial t}+V_{ef}(x)+
\left(\frac{\partial S(x,t)}{\partial x}\right)^2+
Q(x,t)&=&0,\\ \nonumber
& & \\
\frac{\partial R(x,t)}{\partial t}+
2\frac{\partial S(x,t)}{\partial x}\frac{\partial R(x,t)}{\partial x}+
R(x,t)\frac{\partial^2 S(x,t)}{\partial x^2}&=&0\\ \nonumber
\end{eqnarray}
where $V_{ef} = 3k-2\Lambda x$ and the Bohmian quantum 
potential $Q(x,t)$ is defined by \cite{holland},
\begin{equation}
\label{16}
Q(x,t)=-\frac{1}{R(x,t)}\frac{\partial^2 R(x,t)}{\partial x^2}.
\end{equation}

In the present situation, using the value of $R (x,t)$ 
eq. (\ref{13}), $Q (x,t)$ eq. (\ref{16}) takes the form,


\begin{equation}
\label{17}
Q(x,t) = \frac 1 4 \frac{1}{M_1^2} \left( \frac{\partial M_1}{\partial x}\right)^2 - 2 \frac{M_2}{M_1},
	\end{equation}
	
\noindent
where
{\small
\begin{equation}
M_1 = \sum_{n,m=0}^{N} C_n C_m \ Ai\left(G(k,\Lambda, E_n, x)\rule{0mm}{4mm}\right) Ai\left(G(k,\Lambda, E_m, x)\rule{0mm}{4mm}\right) \cos\left((E_n - E_m)t\rule{0mm}{4mm}\right);\end{equation}}
{\small
\begin{eqnarray}
	M_2 &=& \left( 9 \Lambda^2 \right)^{1/3} \sum_{n,m=0}^{N} 
	 C_{{n}}C_{{m}}
	 	 \left[\rule{0mm}{7mm}   \left(G \left( k,\Lambda,E_{{n}},x \right) + G \left( k,\Lambda,E_{{m}},x \right) \rule{0mm}{4mm} \right)
	 	 	\times \right. \nonumber \\
	 	& & \times Ai \left( G \left( k,\Lambda,E_{{n}},x \right)\rule{0mm}{4mm}  \right) 
	 		Ai \left( G \left( k,\Lambda,E_{{m}},x \right) \rule{0mm}{4mm} \right)
	 		  \nonumber \\
	 	& &	\left. + 2\ {Ai}^ {\prime} \left( G \left( k,\Lambda,E_{{n}},x \right) \rule{0mm}{4mm} \right) 
	 		{Ai}^ {\prime} \left( G \left( k,\Lambda,E_{{m}},x \right) \rule{0mm}{4mm} \right)\rule{0mm}{7mm}  \right] \cos\left((E_n - E_m)t\rule{0mm}{4mm}\right)\end{eqnarray}}
	
\noindent $N=20$ and
\begin{equation}
	{Ai}^ {\prime} \left( G \left( k,\Lambda,E_n,x \right) \rule{0mm}{4mm} \right) = \left.\frac{\partial Ai(u)}{\partial u}\right|_{u=G \left( k,\Lambda,E_n,x \right)}.
	\end{equation}


The Bohmian trajectory for $x$ is given by \cite{holland},
\begin{equation}
\label{18}
p_{x}=\frac{\partial S}{\partial x}.
\end{equation}
Computing Hamilton's equation from the superhamiltonian 
eq. (\ref{3}) one notices that $p_x = -6 dx/dt$. If one 
introduces that result in eq. (\ref{18}), one obtains,
\begin{equation}
\label{19}
\frac{dx}{dt}=-\frac{1}{6}\frac{\partial S}{\partial x}.
\end{equation}
Using the value of $S(x,t)$ eq. (\ref{14}) in eq. (\ref{19}), it
reduces to,
%
%
\begin{equation}
\label{20}
	\frac{dx(t)}{dt} = -\left( \frac{\Lambda}{9}\right)^{-1/3} 
	\frac{F_3(t)}{F_4(t)}
	\end{equation}
	
where $N=20$,
{\small
\begin{eqnarray}
	F_3(t) &=& \sum_{n, m=0}^{N}
		C_{n} C_{m}  
		{Ai}^{\prime}\left( G \left( k,\Lambda,E_{{n}},x \left( t \right)\right)\rule{0mm}{4mm}\hspace{-1mm}\right)
		{Ai}\left( G \left( k,\Lambda,E_{{m}},x \left( t \right) \right)\rule{0mm}{4mm}\hspace{-1mm}\right) \times \nonumber \\
		& & \times 
		 \sin \left( \left( E_{{n}}-E_{{m}} \right)t\rule{0mm}{4mm}\right),
	\end{eqnarray}
	
\begin{eqnarray}
	F_4(t) &=& \left[ \sum _{n=0}^{N} C_{n}
		 {Ai}\left( G \left( k,\Lambda,E_{{n}},x \left( t \right)\right)\rule{0mm}{4mm}\hspace{-1mm}\right) 
		\cos \left( E_{n}t \right)  \right] ^{2} + \nonumber \\
		& & 
		 \left[\sum_{m=0}^{N} C_{m}
		 Ai \left( G \left( k,\Lambda,
		E_{{m}},x \left( t \right)\right)\rule{0mm}{4mm}\hspace{-1mm}\right) 
		\sin \left( E_{{m}}t \right)  \right] ^{2}.
	\end{eqnarray}	}

The solution to equation (\ref{20}), which is the Bohmian 
trajectory of $x$, which is the variable describing the
universe, represents the quantum behavior for the cosmic 
evolution in the Planck era.

\begin{figure}[h!]
\includegraphics[width=8cm,height=7cm]{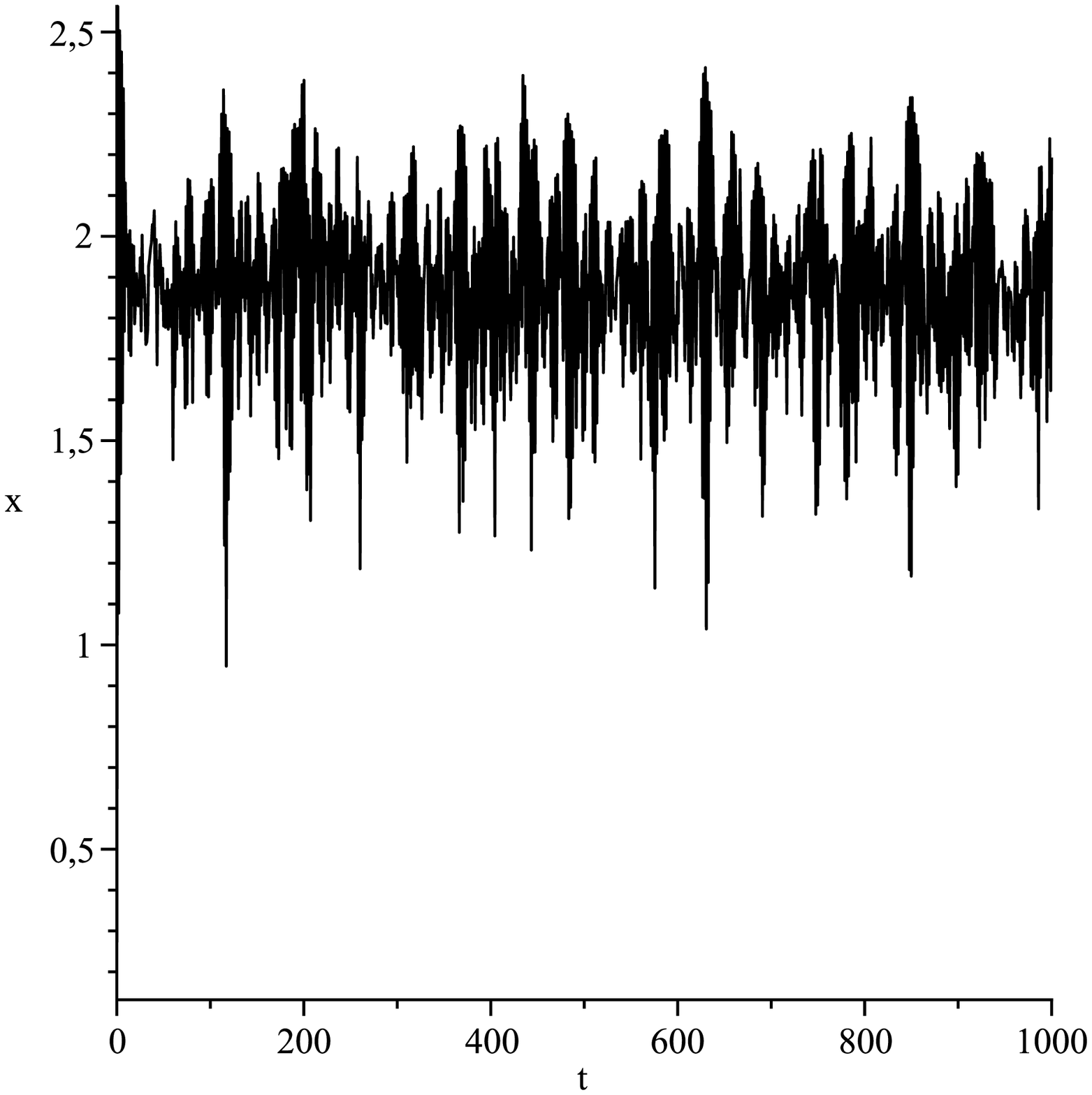}
\caption{Time evolution of $x$ for a FRW model with $k=-1$, $\Lambda=-15$ 
and a perfect fluid of cosmic strings ($p=-1/3\rho$).
The wavefunction was obtained by the linear combination of the twenty-one lowest
energy eigenfunctions.}
\label{f1}
\end{figure}

We have solved eq. (\ref{20}) for several values of negative
$\Lambda$, all possible values of $k$ and many different wavefunctions
constructed as linear combinations of the energy eigenfunctions and
corresponding time sectors. We have found the same qualitative behavior 
for the Bohmian trajectories of $x$, in all those cases. 
It oscillates between a maximum and a minimum value and never goes
through the zero value. It means that, quantum mechanically, in
those models there are no big bang singularities which confirms the
result obtained using the many worlds interpretation \cite{pedram}.
In order to exemplify this behavior we show the Bohmian trajectory
of $x$ for a model with $k = - 1$, $\Lambda = - 15$ and the wavefunction
obtained by the linear combination of twenty-one energy eigenfunctions and
corresponding time sectors with eigenvalues given in Table \ref{tabela1}. 
Those are the twenty-one lowest energy eigenvalues and are very similar to 
those found in \cite{pedram}, specially for large values of $n$. For 
simplicity we have set the twenty-one 
constants coming from the linear combination equal to $C_n = (-1)^{n+1}$,
where $n = 0, 1,..., 20$. We have computed the time evolution of $x$ up to 
$t=1000$ and used the initial condition $x=0.152055822977591$ at $t=0$. 
This initial condition was obtained from the calculation of the expected 
value of $x$, for the same model and the same linear combination of energy 
eigenstates and corresponding time sectors. The result is shown in Figure 
\ref{f1} and is qualitatively similar to the figure representing the scale 
factor expected value given in Ref. \cite{pedram}.

\begin{table}[h!]
\begin{tabular}{|c|c|c|}
\hline
$E_0$ = 6.860180827 & $E_1$ = 14.23955048 & $E_2$ = 20.28110245  \\ \hline 
$E_3$ = 25.62065647 & $E_4$ = 30.50170883 & $E_5$ = 35.04999229  \\ \hline
$E_6$ = 39.34105502 & $E_7$ = 43.42474496 & $E_8$ = 47.33612710  \\ \hline
$E_9$ = 51.10104684 & $E_{10}$ = 54.73924547 & $E_{11}$ = 58.26623169  \\ \hline
$E_{12}$ = 61.69446998 & $E_{13}$ = 65.03416823 & $E_{14}$ = 68.29381974 \\ \hline
$E_{15}$ = 71.48058714 & $E_{16}$ = 74.60058160 & $E_{17}$ = 77.65907027 \\ \hline
$E_{18}$ = 80.66063387 & $E_{19}$ = 83.60928720 & $E_{20}$ = 86.50857447 \\ \hline
\end{tabular}
\caption{The twenty-one lowest energy levels for a FRW model with $k=-1$,
$\Lambda = -15$ and a perfect fluid of cosmic strings ($p=-1/3\rho$).}
\label{tabela1}
\end{table}

\begin{figure}[h!]
\includegraphics[width=8cm,height=7cm]{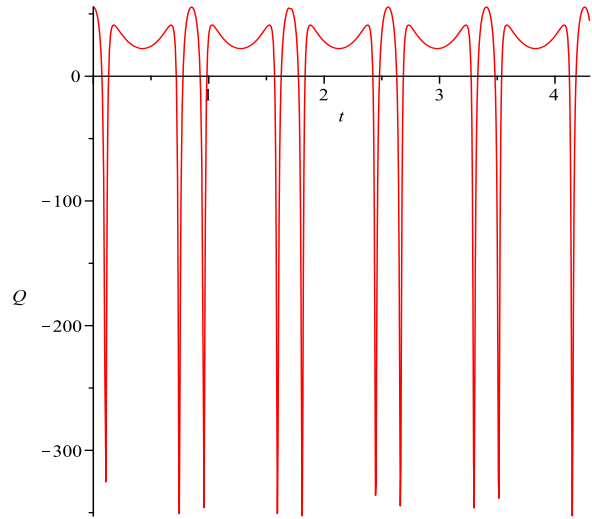}
\caption{Bohmian quantum potential for a FRW model with $k=-1$, $\Lambda=-15$ 
and a perfect fluid of cosmic strings ($p=-1/3\rho$).
The wavefunction was obtained by the linear combination of the two lowest energy 
eigenfunctions.}
\label{f3}
\end{figure}

\begin{figure}[h!]
\includegraphics[width=8cm,height=7cm]{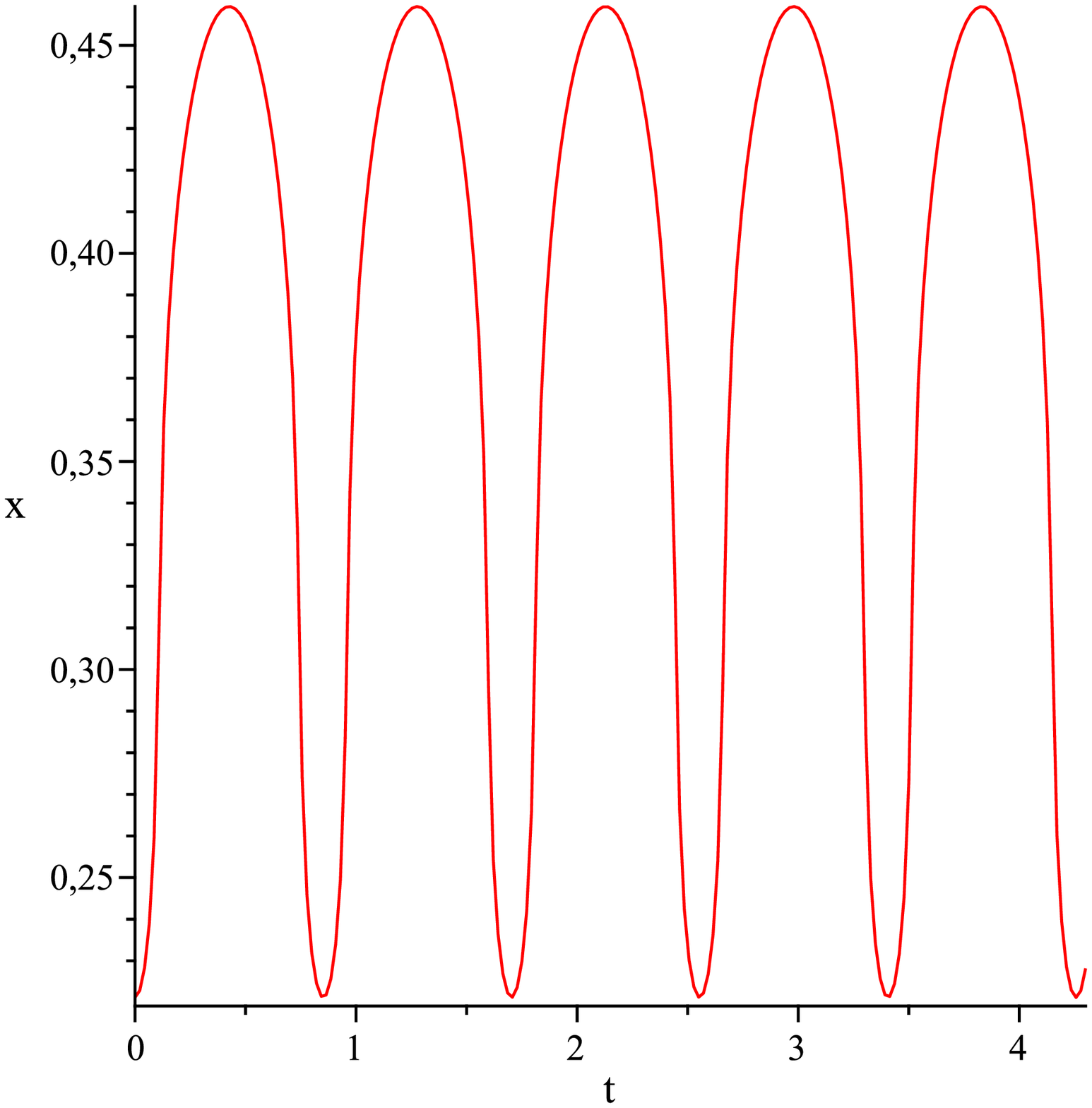}
\caption{Time evolution of $x$ for a FRW model with $k=-1$, $\Lambda=-15$ 
and a perfect fluid of cosmic strings ($p=-1/3\rho$).
The wavefunction was obtained by the linear combination of the two lowest energy 
eigenfunctions.}
\label{f2}
\end{figure}

The absence of big bang singularities in the present models are very easy
to understand when one observes the Bohmian quantum potential eq. (\ref{17}),
for those models. We have computed $Q(x,t)$ eq. (\ref{17}), for several 
values of negative $\Lambda$, all possible values of $k$ and many different 
wavefunctions constructed as linear combinations of the energy eigenfunctions
and corresponding time sectors. 
The calculations were made over the Bohmian trajectories of $x$. It means that
$Q$ reduces to a function that depends only on $t$. We have found the same 
qualitative behavior of $Q$, in all those cases. Initially, at $t=0$, there is
a potential barrier ($B_0$) that prevents the value of $x$ ever to go through zero. 
Then, the barrier becomes a well for a brief moment and again a new barrier 
appears ($B_1$). After a while, $B_1$ turns into a well for a brief moment
and then another barrier identical to $B_0$ appears. After that $Q$,
periodically, repeats itself. $B_0$ is different from $B_1$. $B_1$ exists for a 
longer period and is shorter than $B_0$. One may interpret the potential shape in 
the following way. Initially, at $t=0$, $x$ starts to grow from its minimum value 
different from zero, first rapidly, and then its velocity starts to decrease until 
it goes to zero, at the maximum value of $x$. Then, $x$ starts to decrease, first 
slowly, and then its velocity starts to increase until $x$ reaches its minimum value 
different form zero. There, its velocity 
changes sign and $x$ starts to grow once more, as described above. This dynamics is 
represented in $Q$, initially, by $B_0$, then the first well, then $B_1$ and finally
the well just after $B_1$. Then the movement of $x$ repeats itself periodically.
These models have no big bang singularities because $B_0$ and its periodic repetitions
prevent $x$ ever to go through zero. In order to exemplify this behavior we show, in
Figure \ref{f3}, the Bohmian quantum potential eq. (\ref{17}), for the model with 
$k = - 1$, $\Lambda = - 15$. The wavefunction was obtained by the linear combination of 
two energy eigenfunctions and corresponding time sectors with eigenvalues $E_0$ and 
$E_1$ given in Table \ref{tabela1}. For a better visualization of $Q$'s
behavior, we choose a small time interval in Figure \ref{f3}. For a clearer understand
of $Q$'s behavior we have plotted, in Figure \ref{f2}, the Bohmian trajectory of $x$
for the model with the same conditions described in Figure \ref{f3}, during the same
time interval of Figure \ref{f3} and initial condition $x=0.221185558521407$ at $t=0$. 
This initial condition was obtained from the calculation of the expected 
value of $x$, for the same model and the same linear combination of energy 
eigenstates and corresponding time sectors of the model described in Figure \ref{f3}.




{\bf Acknowledgements.} G. Oliveira-Neto, E. V. Corr\^{e}a Silva and  G. A. 
Mo\-ne\-rat (Researchers of CNPq, Brazil) thank CNPq and FAPERJ
for partial financial support. We thank the opportunity to use the 
Laboratory for Advanced Computation (LCA) of the Department of Mathematics,
Physics and Computation, FAT/UERJ, where part of this work was prepared.

\end{document}